\documentclass[12pt]{article}
\usepackage[utf8]{inputenc}
\usepackage[letterpaper,margin=1in,headsep = 5mm]{geometry}
\usepackage{siunitx}
\usepackage{graphicx}
\usepackage{amsmath}
\usepackage{circuitikz}
\usepackage{setspace}
\usepackage{mathrsfs}
\usepackage[sorting=none]{biblatex}
\usepackage[withpage]{acronym}
\usepackage{authblk}
\usepackage[automake,nonumberlist]{glossaries}
\makeglossaries

\usetikzlibrary{shapes,arrows,calc,positioning,backgrounds,fit,patterns}

\newacronym{ep}{EP}{Electropolishing}
\newacronym{srf}{SRF}{Superconducting Radiofrequency}
\newacronym{hf}{HF}{Hydrofluoric}
\newacronym{eis}{EIS}{Electrochemical Impedance Spectroscopy}
\newacronym{drt}{DRT}{Distribution of Relaxation Times}
\newacronym{nb}{Nb}{Niobium}
\newacronym{gdrt}{gDRT}{Generalized Distribution of Relaxation Times}
\newacronym{v}{V}{Volt}
\newacronym{hz}{Hz}{Hertz}
\newacronym{RC}{RC}{Resistor-Capacitor}
\newacronym{RLC}{RLC}{Resistor-Inductor-Capacitor}
\newacronym{RMS}{RMS}{Root Mean Square}

\newglossaryentry{j}{name={$j$}, description={The Imaginary Unit $\sqrt{-1}$}}
\newglossaryentry{Z}{name={$Z$}, description={Complex Electrical Impedance}}
\newglossaryentry{s}{name={$s$}, description={Complex Frequency}}
\newglossaryentry{R}{name={$R$}, description={Resistance}}
\newglossaryentry{C}{name={$C$}, description={Capacitance}}
\newglossaryentry{L}{name={$L$}, description={Inductance}}
\newglossaryentry{Omega}{name={$ \Omega $}, description={Frequency Scaled Resistance}}
\newglossaryentry{lambda}{name={$ \lambda $}, description={Eigenvalue/Eigenfrequency of a Linear Dynamical System}}
\newglossaryentry{alpha}{name={$ \alpha $}, description={Real Part of a Complex Eigenvalue}}
\newglossaryentry{beta}{name={$ \beta $}, description={Imaginary Part of a Complex Eigenvalue}}
\newglossaryentry{G}{name={$G$}, description={The DRT}}
\newglossaryentry{gamma}{name={$ \gamma $}, description={The DRT in the Log Coordinate System}}
\newglossaryentry{phi}{name={$ \phi $}, description={A Test Function}}
\newglossaryentry{omega}{name={$ \omega $}, description={Angular Frequency}}
\newglossaryentry{delta}{name={$ \delta $}, description={The Dirac Delta Function}}
\newglossaryentry{S}{name={$S$}, description={A Chemical State}}
\newglossaryentry{k}{name={$k$}, description={A Reaction Rate Constant}}
\newglossaryentry{P}{name={$\vec{P}$}, description={A State Vector}}
\newglossaryentry{I}{name={$I$}, description={Electrode Current}}
\newglossaryentry{E}{name={$E$}, description={Electrode Potential}}
\newglossaryentry{Lambda}{name={$\mathbf{\Lambda}$}, description={A Diagonal Matrix}}
\newglossaryentry{J}{name={$\mathbf{J}$}, description={A Jordan Matrix}}
\newglossaryentry{N}{name={$\mathbf{N}$}, description={Upper Shift Matrix}}

\glsaddall

\begin{document}

\title{Analysis of Niobium Electropolishing Using a Generalized Distribution of Relaxation Times Method}
\author[1,2]{Eric Viklund}%
\author[2]{Vijay Chouhan}
\author[2]{Davida Smith}
\author[2]{Tim Ring}
\author[1]{David N. Seidman}
\author[2]{Sam Posen}
\affil[1]{Department of Materials Science and Engineering, Northwestern University}
\affil[2]{Fermi National Accelerator Laboratory}

\date{\today}
\maketitle
\begin{abstract}
  Using electrochemical impedance spectroscopy, we have devised a method of sensing the microscopic surface conditions on the surface of niobium as it is undergoing an electrochemical polishing (EP) treatment. The method uses electrochemical impedance spectroscopy (EIS) to gather information on the surface state of the electrode without disrupting the polishing reaction. The EIS data is analyzed using a so-called distribution of relaxation times (DRT) method. Using DRT, the EIS data can be deconvolved into discrete relaxation time peaks without any a priori knowledge of the electrode dynamics. By analyzing the relaxation time peaks, we are able to distinguish two distinct modes of the EP reaction. As the polishing voltage is increased, the electrode transitions from the low voltage EP mode, characterized by a single relaxation time peaks, to the high voltage EP mode, characterized by two relaxation time peaks. We theorize that this second peak is caused by the formation of an oxide layer on the electrode. We also find that this oxide induced peak transitions from to a negative relaxation time, which is indicative of a blocking electrode process. By analyzing EPed samples, we show that samples polished in the low voltage mode have significantly higher surface roughness due to grain etching and faceting. We find that the surface roughness of the samples only improves when the oxide film peak is present and in the negative relaxation time region. This shows that EIS combined with DRT analysis can be used to predict etching on EPed Nb. This method can also be performed before or during the EP, which could allow for adjustment of polishing parameters to guarantee a smooth cavity surface finish.
\end{abstract}

\newpage

\printglossary[title={List of Terms}]

\newpage

\section{Introduction}
\label{sec:org5ef967f}
Electropolishing (EP) is commonly the method of choice to polish Nb superconducting radiofrequency (SRF) cavities. EP is an electrochemical polishing method using an HF-containing electrolyte to dissolve the surface of the Nb under an applied votage. In ideal conditions, EP can acheive nanometer-scale surface roughness. However, defects such as etching and pitting can sometimes occur, especially when polishing large cavities. One of the main causes of these defects has been shown to be insufficient voltage applied to the cavity during EP.\cite{chouhan2022study,viklund2022studies,chouhan2023electropolishing} To better understand why this is the case, we utilize a technique known as electrochemical impedance spectroscopy (EIS) to investigate the surface chemistry of Nb in the HF electrolyte.

EIS has been used to study the chemistry of niobium EP previously.\cite{cattarin2002nb,Tian_2008, tian2008novel, ranjith2018anodic} However, these studies only provide a basic analysis of the chemistry and rely on pre-supposed equivalent circuits to explain the measured impedance spectrum.

The goal of this study is to utilize EIS measurements to find the chemical mechanism behind the etching phenomenon without making apriori assumptions such as relying on a circuit model. To accomplish this, we use a model-free method of analyzing the EIS spectrum called distribution of relaxation times (DRT) analysis.\cite{10.1063/1.1745355, wan2015influence, ZHANG2015464} DRT analysis is used to extract the relaxation times of the chemical processes occuring on the electrode without the use of an equivalent circuit model. This method has been used to characterize a wide range of electrochemical systems such as fuel cells,\cite{Sonn_2008, schichlein2002deconvolution, Leonide_2008} batteries,\cite{SCHMIDT201370, batteries5020043, SONI202297} and supercapacitors.\cite{HELSETH2019100912} In section \ref{sec:org7d749e2} we show that these relaxation times are equivalent to the eigenvalues of the linearized electrochemical system dynamics. With this new insight we generalize the DRT method to negative eigenvalues which correspond to unstable system dynamics, which is a necessary step for understanding the impedance characteristics of Nb EP.

Using this generalized DRT method, we are able to observe the formation of an NbO and an Nb\textsubscript{2}O\textsubscript{5} layer on the Nb electrode as the polishing voltage is increased. The formation of the Nb\textsubscript{2}O\textsubscript{5} oxide also corresponds with a reduction in surface etching indicating that the formation of this oxide layer is necessary to acheive good polishing.

\subsection{Niobium Electropolishing Using HF-containing Electrolytes}

Electropolishing (EP) is the main method used to polish the surface of Nb SRF cavities. EP is neccessary to achieve a smooth surface finish and to remove the damaged layer caused by the shaping and welding of the Nb sheets during cavity manufacturing. A proper EP can reduce the RMS surface roughness of the cavity below \qty{20}{\nano\meter} and is essential for achieving good cavity performance.

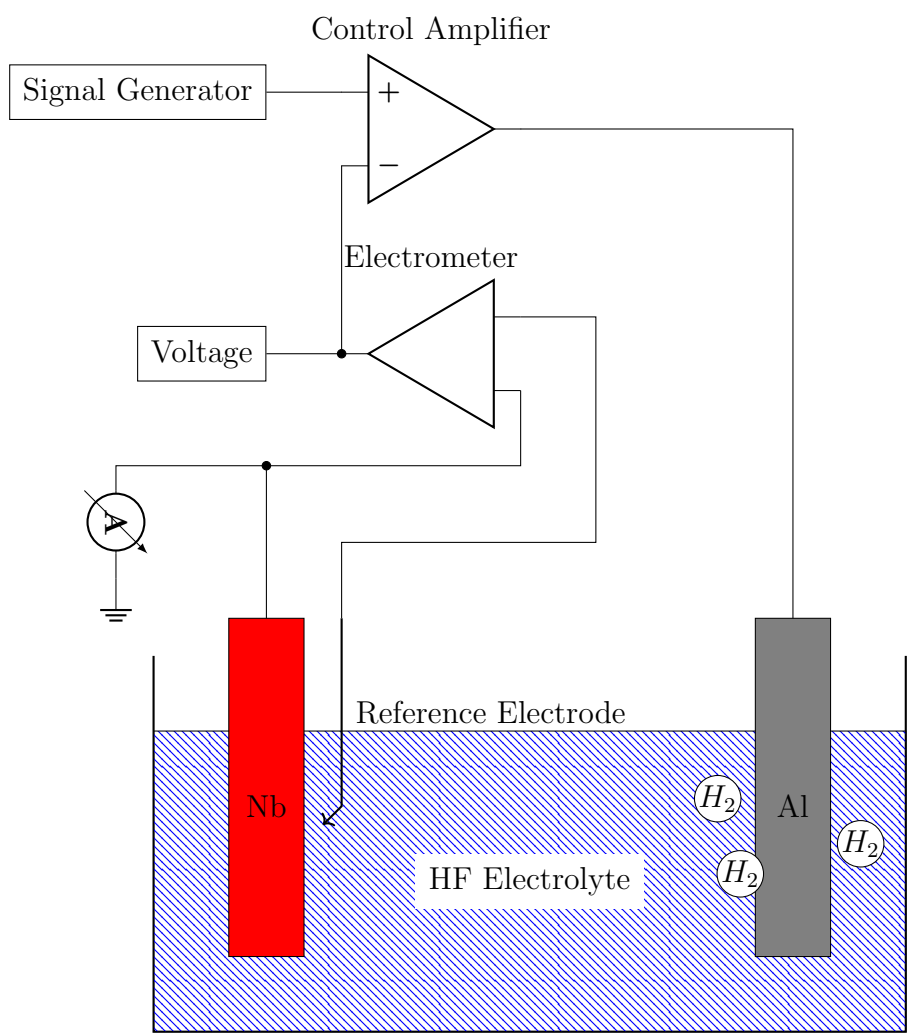
\begin{figure}[htbp]
    \centering
    
    \begin{circuitikz}[inner sep=0pt,
                    every label/.style={minimum size=6mm}]
    \draw[thick] (0.0,5.0) -- (0.0,0.0) -- (10.0,0.0) -- (10.0,5.0);
    \draw[pattern=north west lines, pattern color=blue] (0,0) rectangle (10.0,4.0);
        \node[fill=white,inner sep=5pt] at (5.0,2.0) {HF Electrolyte};

    \draw[fill=red] (1.0,1.0) rectangle (2.0,5.5);
        \node at (1.5,3.0) {Nb};
        \coordinate (WE) at (1.5,5.5);

    \draw[fill=gray] (9.0,1.0) rectangle (8.0,5.5);
        \node at (8.5,3.0) {Al};
        \coordinate (CE) at (8.5,5.5);
        \node[shape=circle,fill=white,draw] at (7.5,3.1) {$H_2$};
        \node[shape=circle,fill=white,draw] at (9.4,2.5) {$H_2$};
        \node[shape=circle,fill=white,draw] at (7.8,2.1) {$H_2$};

    \draw[thick,->] (2.5, 5.5) coordinate(Ref) to node[right,pos=0.5,inner sep=5pt]{Reference Electrode} (2.5, 3.0) 
        -- ++(-0.25,-0.25); 

    \draw (WE) ++(0,7) node[left,inner sep=5pt,draw]{Signal Generator} to[short] ++(1,0)
        node[op amp, noinv input up, anchor=+, label={[above,label distance=10mm] Control Amplifier}](Control Amp){}
        (Control Amp.-) -- ++(0,-2.5) coordinate(Vout)
        to[short, *-] ++(-1,0)
        node[left,inner sep=5pt,draw]{Voltage}
        (Control Amp.out) to (Control Amp.out -| CE) -- (CE)
        ;

        \draw (Vout) to ++(0,0) 
        node[plain amp, xscale=-1, anchor=out, label={[above,label distance=10mm] Electrometer}](Electrometer){}
        (Electrometer.in up) to ++(1,0) -- ++(0,-3) coordinate(Refup) -- (Refup -| Ref) -- (Ref)
        (Electrometer.in down) to ++(0,-1) coordinate(WEup) -- (WEup -| WE) to[short,*-] (WE)
        ;

        \draw (WEup -| WE) to ++(-2,0) to[ammeter] ++(0,-1.5) node[ground]{}
        ;

\end{circuitikz}
    \caption{A schematic of a Nb electropolishing cell showing the potentiostat circuit diagram and the connection to the Nb EP cell.}
    \label{fig:EP_diagram}
\end{figure}

Niobium is part of a group of metals known as valve metals, group IV and V elements on the periodic table that are highly resistant to chemicals. Other metals in this group include Zirconium, Tantalum, and Hafnium. All of these metals are characterized by the formation of a highly stable, barrier type oxide film on the surface under anodic conditions. This stable oxide protects the metal from reacting with most chemicals. 

To use EP on Nb SRF cavities, a chemical that can remove the oxide layer is neccessary. Hydroflouric (HF) acid is one of the few chemicals that can dissolve this stable oxide film. The niobium is submerged in an electrolyte consisting of HF and H\textsubscript{2}SO\textsubscript{4} and a positive voltage is applied causing the surface to oxidize. The HF in the electrolyte can then react with the oxide layer resulting in material removal. The reaction is balanced with an aluminum counter electrode which reacts via simple hydrolysis. 

The reaction of Nb with the HF electrolyte changes as the applied voltage changes. At low voltages, the Nb is in the active region, which is characterized by a positive correlation between the anodic current and the applied voltage. In this region, the dissolution reaction is limited by the oxidation rate of Nb into its oxide form, which is dependant on the applied voltage. As the voltage increases, the current reaches a maximum point after which the reaction is said to be in the passive region, characterized by an anodic current plateau where the current stays constant at all higher voltages. In this region, the reaction is limited by the rate of oxide dissolution by HF. The dissolution of the oxide occurs without any charge transfer, which means the reaction rate only depends on the concentration of HF in the electrolyte and is independant of the applied voltage\cite{ranjith2018anodic}.

Depending on the applied voltage, different Nb oxides may exist on the surface. The most common oxide is Nb\textsubscript{2}O\textsubscript{5} which is the oxide present on the surface when the Nb is passivated. This oxide can only be dissolved via a chemical reaction with undissociated HF molecules in the electrolyte. The rate of dissolution is determined by the HF concentration in the electrolyte as well as the diffusion rate of HF to the surface of the electrode. This diffusion rate can be increased by mixing the electrolyte reducing the thickness of the HF depletion layer near the electrode\cite{tian2010evaluation}. At lower voltages Nb may be oxidized to NbO instead. This oxide has additional electrochemical pathways for dissolution by reacting with $HF_2^-$ ions which increases the anodic current at low voltages leading to a peak in the current-voltage curve followed by a negative current-voltage relation\cite{ranjith2018anodic}. This negative relation can cause instabilities leading to spontaneous current oscillations which are commonly observed during cavity EP\cite{khun2013smoothening}.

\subsection{Electrochemical Impedance Spectroscopy}

Electrochemical impedance spectroscopy (EIS) is a measurement technique used to extract information about the electrode reaction mechanism. The measurement is performed by applying a small voltage or current perturbation at different frequencies to the electrode and measuring the response. This gives the complex impedance of the electrode as a function of the perturbation frequency. The impedance spectrum must then be interpreted to gain useful information.

EIS measurements are performed using a potentiostat in combination with a sinusoidal signal generator. A potentiostat is an instrument used to maintain a given potential on the working electrode, in this case the Nb electrode. This is done using a reference electrode, a third electrode placed close to the working electrode, which measures the electric potential in the electrolyte near the working electrode. Using this measurement, the ohmic losses in the electrolyte as well as losses in the counter electrode can be compensated. When a sinusoidal signal is sent to the potentiostat, the working electrode potential will also oscillate sinusoidally. The electrode current can then be measured to calculate the complex impedance. The potentiostat circuit diagram is shown in figure~\ref{fig:EP_diagram}.

There are two main conditions required for EIS measurements to be valid. The first is that the electrode process is stationary, meaning that state of the system stays constant thorughout the duration of the experiment. This does not mean that the system is in thermodynamic equillibrium, since the reaction may progress at a constant rate and still uphold the stationary condition. The second condition is that the electrochemical system must be linear with respect to the applied perturbation. This means that the perturbation must be small enough that the current-voltage relation can be approximated as linear, since electrochemical systems are rarely linear in practice. Typically, a perturbation between \qtyrange{10}{100}{\milli\volt} is used.

Following the principles of linear systems thoery, the impedance of an electrochemical system can be represented by a series of relaxation processes where $\frac{1}{\lambda}$ is the relaxation time and $\Omega$ is the impedance contribution from the relaxation process. $R$ is the resistance stemming from ohmic losses in the system such as electrolyte resistance.

\begin{flalign}
    Z\left(s\right) &= R + \sum_{i}\frac{\Omega_i}{s-\lambda_i} \label{eq:EIS_damped}
\end{flalign}

This equation is the impedance of a first order linear system, a system that can be described by a set of independant first order differential equations. The electrochemical system may also exhibit second order dynamics characterized by a complex eigenvalue. This can cause effects such as ringing, transient oscillations that decay over time, in the current response when the a step voltage is applied. In this case the impedance has the form of a series of oscillators with a damping factor $\beta$ and a resonant frequency $\alpha$ corresponding to the real and imaginary parts of the eigenvalues.

\begin{flalign}
    Z\left(s\right) &= R + \sum_{i}\frac{\Omega_{re}\left(s-\alpha_i\right)+\Omega_{im}j\beta_i}{s^2-2s\alpha_i+\alpha_i^2+\beta_i^2} \label{eq:EIS_underdamped}
\end{flalign}

\begin{figure}[htbp]
    \centering
    
    \includegraphics[width=\textwidth]{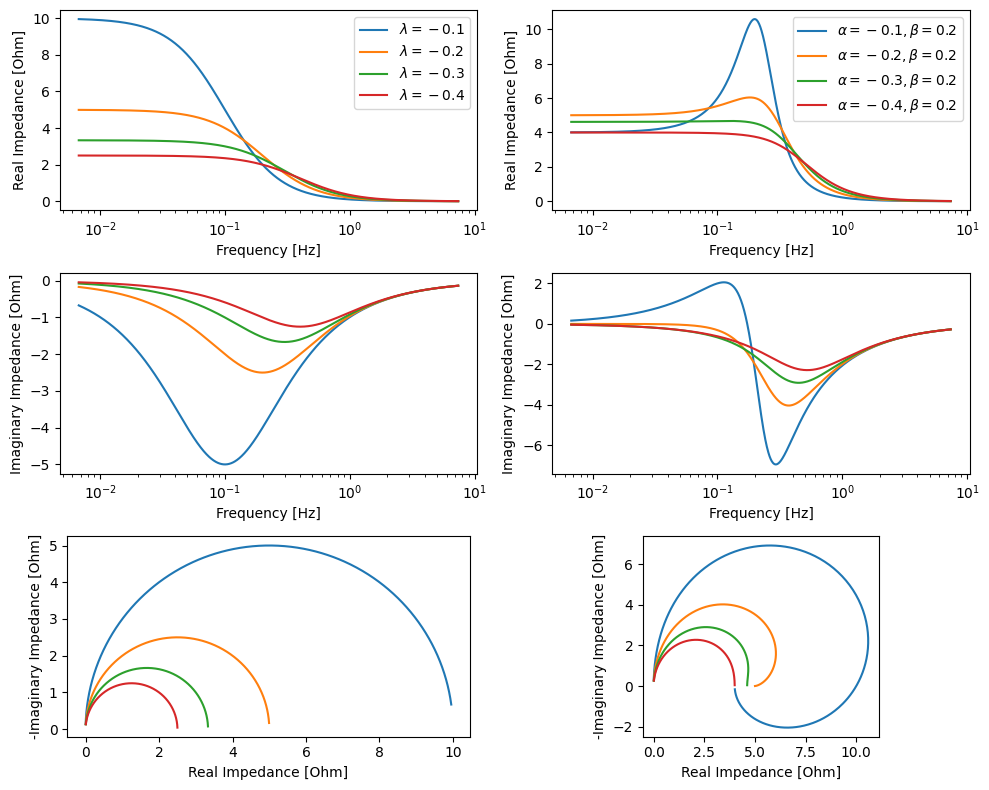}
    \caption{The characteristic impedance spectra of first order dynamics according to equation~\ref{eq:EIS_damped} (left) and second order dynamics according to equations~\ref{eq:EIS_underdamped} (right). $\Omega = 1$ for all plots.}
    \label{fig:EIS_plots}
\end{figure}

See section~\ref{sec:sup} for the derivation of equations \ref{eq:EIS_damped} and \ref{eq:EIS_underdamped} as well as other higher order forms of the impedance response. The impedance of these two types of system is plotted in figure~\ref{fig:EIS_plots}. The real impedance of the damped system approaches a maximum value at zero frequency whereas the underdamped system reaches a maximum real impedance at the resonant frequency causing a resonance peak. This peak is essential for distinguishing between the two types of dynamics in when analyzing EIS data. By fitting the experimental impedance measurements to these equations, the time scales of the different reaction mechanism on the electrode can be determined. Ultimately, the goal of performing EIS measurements on Nb EP is to measure the impedance over a range of voltages and polishing conditions to find the mechanisms that are responsible for good electropolishing results and those responsible for poor results such as etching or pitting.

\subsection{Generalized Distribution of Relaxation Times}
\label{sec:org7d749e2}

As shown previously, the impedance spectrum is related to the eigenvalues of the electrochemical system's linearized dynamics. To calculate these eigenvalues from the experimental EIS measurements, we use the distribution of relaxation times (DRT) method. DRT is a flexible and general method that can be used to model impedance spectra without the use of an equivalent circuit model. The benefit of this method is that it does not require a presupposed circuit model, which may not be correct or have any physical meaning. The DRT method also allows for a linear fitting algorithm to be used instead of a non-linear one as shown in section~\ref{sec:sup}.

DRT analysis is typically based on an assumption that the electrode reaction can be modeled by a series of voigt circuits, a resistor and a capacitor connected in parallel. The electrode is simulated as a string of circuit elements consisting of infinitesimal voigt elements which can describe arbitrary capacitive impedance spectra. The impedance of Nb EP exhibits characteristics that cannot be described using only resistors and capacitors as discussed in section~\ref{sec:results}. Therefore, it is necessary to generalize the DRT method to include these features.

We can express the impedance of the system as an integral equation where $G$ is the relaxation time distribution, $\omega_0$ is the inverse of the relaxation time also known as the natural frequency.

\begin{flalign}
  Z_{gDRT}\left(j \omega\right) &= \int_{-\infty}^{\infty} \frac{G\left(\omega_0\right) d\omega_0}{1 + j\frac{\omega}{\omega_0}} \label{eq:gDRT}
\end{flalign}

It was previously shown that the impedance of a linear system with first order dynamics can be written as a sum of the impedance contribution from each of the system's eigenvalues. The previous integral equation can be reduced to this form if $G$ is a sum of Dirac delta functions.

\begin{flalign}
  G\left(\omega_0\right) &= \sum_{n=1}^{N}\frac{\Omega_i}{\omega_0}\delta\left(\omega_0+\lambda_i\right)\\
  Z_{gDRT}\left(j \omega\right) &= \int_{-\infty}^{\infty} \frac{\sum_{n=1}^{N}\frac{\Omega_i}{\omega_0}\delta\left(\omega_0+\lambda_i\right) d\omega_0}{1 + j\frac{\omega}{\omega_0}}\\
  Z_{gDRT}\left(j \omega\right) &= \sum_{n=1}^{N} \frac{\Omega_{i}}{j \omega - \lambda_{i}} \\
\end{flalign}

From this we can see that $G$ and $\omega_0$ are related to $\Omega$ and $\lambda$. For a real measurement, the peaks in the DRT will not be perfect delta functions, but instead will have some width due to measurement noise and the finite number of data points. For a peak centered on the frequency $\omega_i$ with a width of $2a$ we can approximate $\Omega_i$ and $\lambda_i$ as

\begin{flalign}
  \Omega_i &\approx \int_{\omega_i-a}^{\omega_i+a}\omega_0 G\left(\omega_0\right) d\omega_0\\
  \lambda_i &= -\omega_i \\
\end{flalign}

The eigenvalues of the electrochemical system are therefore related to the peaks in the distribution time. The frequency of those peaks gives the eigenvalues of the system. Positive frequencies correspond to stable eigenvalues, $\lambda < 0$, and negative frequencies correspond to unstable eigenvalues, $\lambda > 0$. The area under each peak is proprtional to the contribution of each eigenvalue to the total impedance spectrum. DRT can in principle also model higher order impedance spectra, which are present in systems with complex or repeated eigenvalues. However, in this case the DRT will not take the form of delta functions. A non-homogeneous electrode surface may also affect the DRT. Rather than a single eigenvalue, a non-homogeneous electrode will have a distribution of eigenvalues caused by the varying surface conditions. This effect would cause the sharp peaks of the DRT to become more broad.

\section{Experimental}
\label{sec:orgb71f960}

The electrolyte used for this experiment is composed of a 9:1 ratio of 98\% by volume sulfuric acid and 70\% by volume hydroflouric acid. A strip of Nb foil was prepared by cleaning with alcohol. Approximately \qty{8.4}{\centi\meter\squared} surface area of the strip was exposed to the electrolyte. A Nb wire was used as a reference electrode, since it is resistant to the EP electrolyte. The counter electrode is an aluminum rod. The temperature of the electrolyte was held at \qty{21}{\celsius} using an aluminum cooling coil immersed in the electrolyte. The EIS measurements were performed using a BioLogic VSP-300 potentiostat over a frequency range of \qtyrange{0.5}{2e5}{\hertz}. The EIS measurement was repeated for polishing voltages ranging from \qtyrange{0.5}{0.9}{\volt} measured versus the open circuit potential.

A series of Nb samples were electropolished using voltages ranging from \qtyrange{0.5}{1.2}{\volt} to determine the effect of the polishing voltage on the surface finish of the Nb. To remove any initial surface roughness from the samples a standard EP treatment was applied. \qty{10}{\micro\meter} of material was removed at a voltage of \qty{16}{\volt} creating a smooth surface finish. Then, \qty{5}{\micro\meter} of material was removed from each sample at lower voltages. To measure the surface finish after the low voltage polishing we use a Keyence VX-K 3000 confocal laser microscope to measure the 2D surface profile of the samples.

\section{Results}
\label{sec:results}

\begin{figure}
  
  \includegraphics[width=\textwidth]{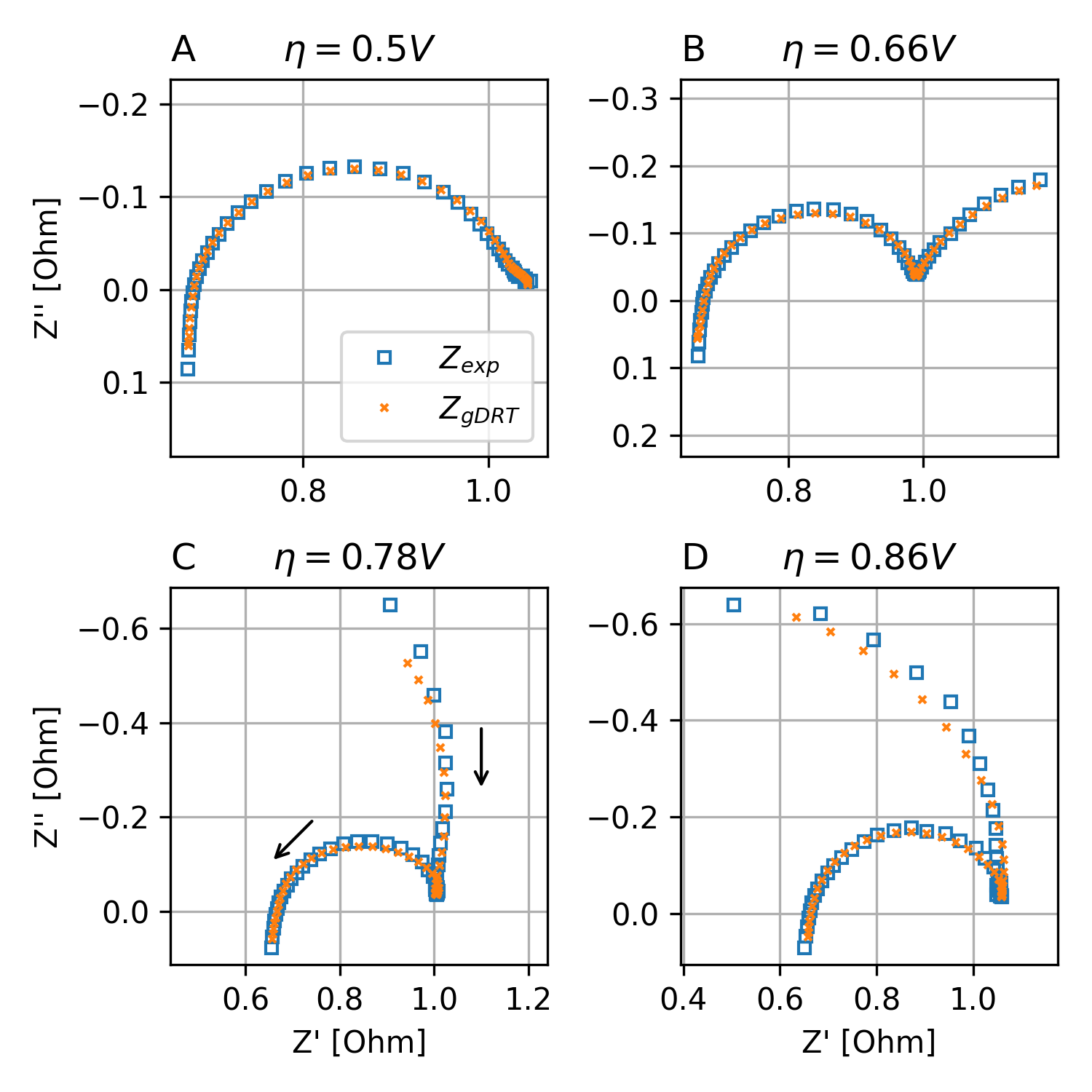}
  \caption{The complex impedance of Nb. Each plot shows the different regimes of electropolishing at low voltage. A one capacitive loop, B two capacitive loops, C capacitive, inductive, and capacitive loops, D capacitive, inductive, and capacitive loop with a negative resistance. The arrows indicate the direction of increasing frequency.}
  \label{fig:nyquistplot}
\end{figure}

EIS measurements reveal a complex evolution of the surface chemistry as the polishing voltage is increased through the \qtyrange{0.5}{1.2}{\volt} range. As shown in figure~\ref{fig:nyquistplot}, at \qty{0.5}{\volt} the system exhibits a single capacitive impedance loop. This is most likely caused by the polarization of the electrode caused by the formation of an electrical double-layer, a layer of adsorbed ions on the electrode surface.\cite{eliaz2018physical} At these low voltages the reaction is charge transfer limited; the transfer of charged ions from the electrolyte to the Nb surface through the double-layer to oxidize the surface is the limiting reaction step.

When the voltage is increased two new features appear, an inductive impedance loop at mid-frequencies and a second capacitive impedance loop at low frequencies. This low frequency loop is caused by the formation of an oxide film and depletion of the HF near the electrode surface. This limits the current through the electrode at higher voltages. When the voltage is increased from \qty{0.78}{\volt} to \qty{0.86}{\volt} the direction of the low frequency loop changes from counter-clockwise to clockwise as shown in figure~\ref{fig:nyquistplot} indicating that the process dynamics have become unstable.

\subsection{Distribution of Relaxation Times of Niobium Electropolishing}

\begin{figure}
  
  \includegraphics[width=\textwidth]{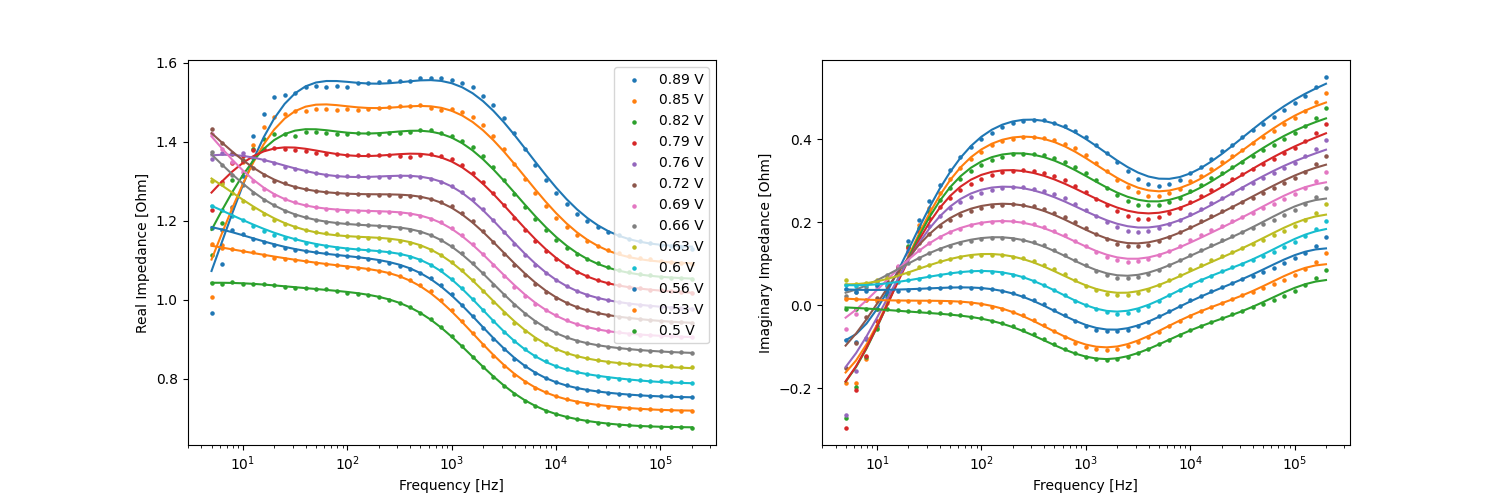}
  \caption{The complex impedance of Nb in electropolishing electrolyte measured using EIS. Each curve is offset from zero. The solid line shows the simulated impedance spectrum from the DRT model.}
  \label{fig:bodeplot}
\end{figure}

\begin{figure}
  
  \includegraphics[width=\textwidth]{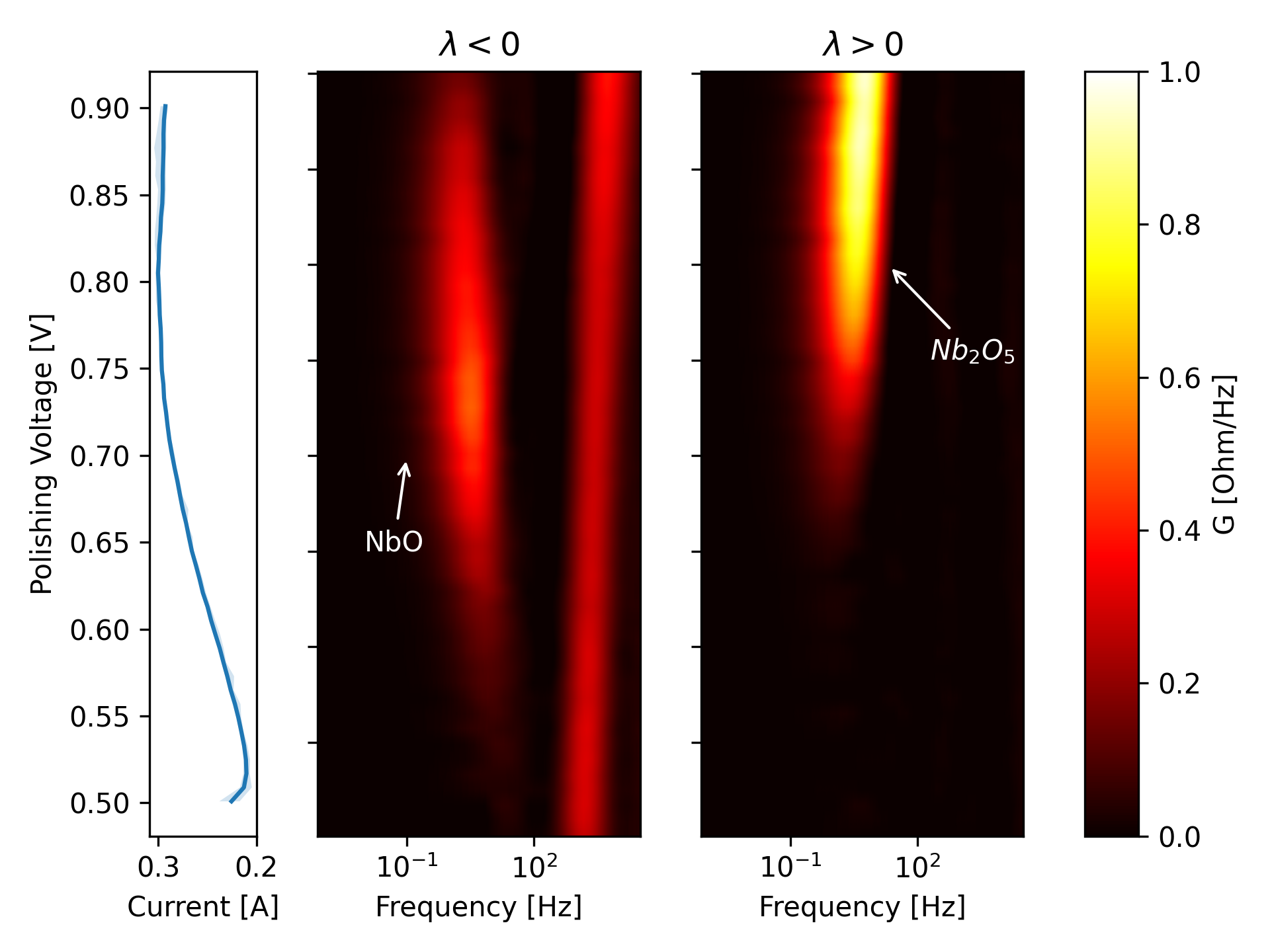}  
  \caption{The Distribution of relaxation times calculated from the EIS impedance data for each of the measured potentials. The current-voltage curve is shown in the graph on the left. The left DRT plot shows eigenvalues that are less than zero, which correspond to stable eigenvalues. The right DRT plot shows eigenvalues that are greater than zero, which correspond to unstable eigenvalues.}
  \label{fig:gamma}
\end{figure}

Using the generalized DRT method, we are able to fit the Nb impedance spectrum across all frequencies including the inductive mid-frequency loop and the low frequency loop as seen in figure \ref{fig:bodeplot}. From figure \ref{fig:gamma} we can see that there are three prominent peaks in the DRT. The peak centered around \qtyrange{100}{1000}{\hertz} corresponds to the electric double layer capacitance. This is clear because the peak exists at all voltages including low voltage. The center of this peak shifts to the right linearly with the polishing voltage. The frequency is plotted on a logarithmic scale, which shows that the relaxation time of the electric double layer follows an Arrhenius type relation.

The two other peaks centered around \qty{10}{\hertz}, one with a stable eigenvalue and one with an unstable eigenvalue. These peaks likely correspond to niobium oxidation reactions. The stable peak with a maximum around \qtyrange{0.7}{0.75}{\volt} is caused by the oxidation of Nb to NbO, while the unstable peak with a maximum at \qtyrange{0.8}{0.9}{\volt} is caused by the oxidation of NbO to Nb\textsubscript{2}O\textsubscript{5}. This can be verified by XPS measurements of Nb exposed to HF electrolytes in the active and passive regimes.\cite{ranjith2018anodic} The NbO peak first appears in the active region of the current voltage curve and reaches a maximum when the Nb transitions to the passive region. The Nb\textsubscript{2}O\textsubscript{5} peak appears in the active to passive transition region and grows larger in the passive region. This indicates that the Nb surface is becoming more oxidized as the polishing voltage increases and more of the surface is being covered by Nb\textsubscript{2}O\textsubscript{5}.

\section{Discussion}

\begin{figure}
  \centering
  
  \begin{tikzpicture}[inner sep=0pt,minimum size=12mm,
                    every label/.style={minimum size=6mm}]
    \node at (-1.5, 0.0) (Nb)    [circle,draw,label=left:$S_0$]         {$Nb$};
    \node at ( 1.0, 2.0) (NbO)   [circle,draw,label=above:$S_1$]        {$NbO$}
        edge [->, bend left = 10]   node [auto, minimum size=1mm]       {$k_{10}$}  (Nb)
        edge [<-, bend right = 10]  node [auto, swap, minimum size=1mm] {$k_{01}$}  (Nb);
    \node at ( 1.0,-2.0) (Nb2O5) [circle,draw,label=below:$S_2$]        {$Nb_2O_5$}
        edge [->, bend left = 10]   node [auto, minimum size=1mm]       {$k_{21}$}  (NbO)
        edge [<-, bend right = 10]  node [auto, swap, minimum size=1mm] {$k_{12}$}  (NbO);
    \node at ( 4.0, 0.0) (Nbaq)  [circle,draw,label=right:$S_3$]        {$Nb_{sol}$}
        edge [<-]  node [auto, swap, minimum size=1mm]                  {$k_{13}$}  (NbO)
        edge [<-]  node [auto, minimum size=1mm]                        {$k_{23}$}  (Nb2O5);

    \draw [fill=blue!30] ( 3.7  , 3.9) circle (0.25);
    \draw [fill=blue!30] ( 4.3  , 1.6) circle (0.25);
    \draw [fill=blue!30] ( 3.4  , -1.4) circle (0.25);
    \draw [fill=blue!30] ( 4.9  , -2.9) circle (0.25);
    \draw [fill=blue!30] ( 3.2  , -3.5) circle (0.25);

    \begin{scope}[on background layer]
        \draw [fill=black!30] (-3  ,-4.5) rectangle ( 0  , 4.5);
        \draw [fill=blue!30]  ( 0  ,-4.5) rectangle ( 2  , 4.5);
    \end{scope}
\end{tikzpicture}
  \caption{Proposed mechanism for niobium electropolishing in HF-containing electrolyte by a two step oxidation reaction.}
  \label{fig:diagram}
\end{figure}
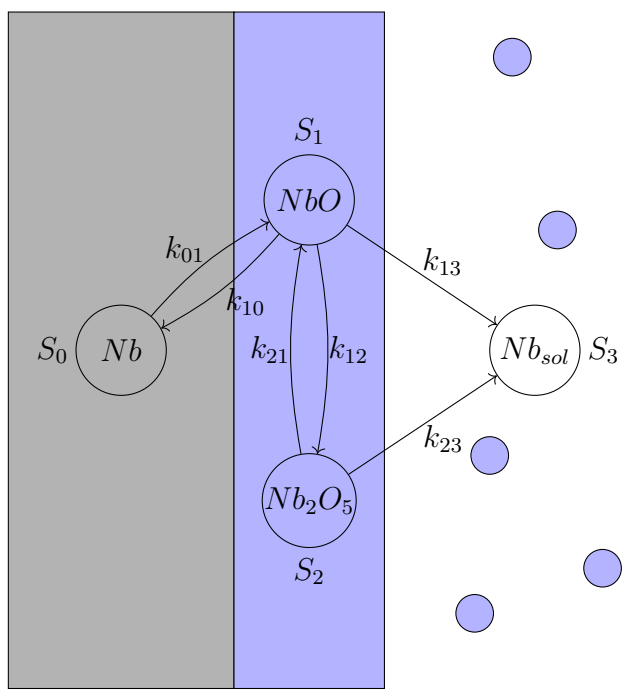

We propose a possible two step oxidation reaction mechanism to explain this behavior. At low voltages, the Nb is oxidized to the $Nb^{2+}$ oxidation state creating NbO. This NbO is immediately dissolved by HF since the reaction rate is limited by the oxidation step at low voltage. The rate of oxidation is dependent on the Nb grain orientation which leads to the observed grain etching phenomenon at low voltage. Once the voltage is increased, the oxidation rate is large enough for a thin layer of NbO to form on the surface as shown by the first oxidation peak in the DRT. As the voltage increases further, the oxidation state of Nb gradually changes to $Nb^{5+}$ through the formation of Nb\textsubscript{2}O\textsubscript{5} as shown by the second oxidation peak in the DRT. A diagram of this reaction mechanism is shown in figure~\ref{fig:diagram}.

\begin{figure}
    
  \includegraphics[width=\textwidth]{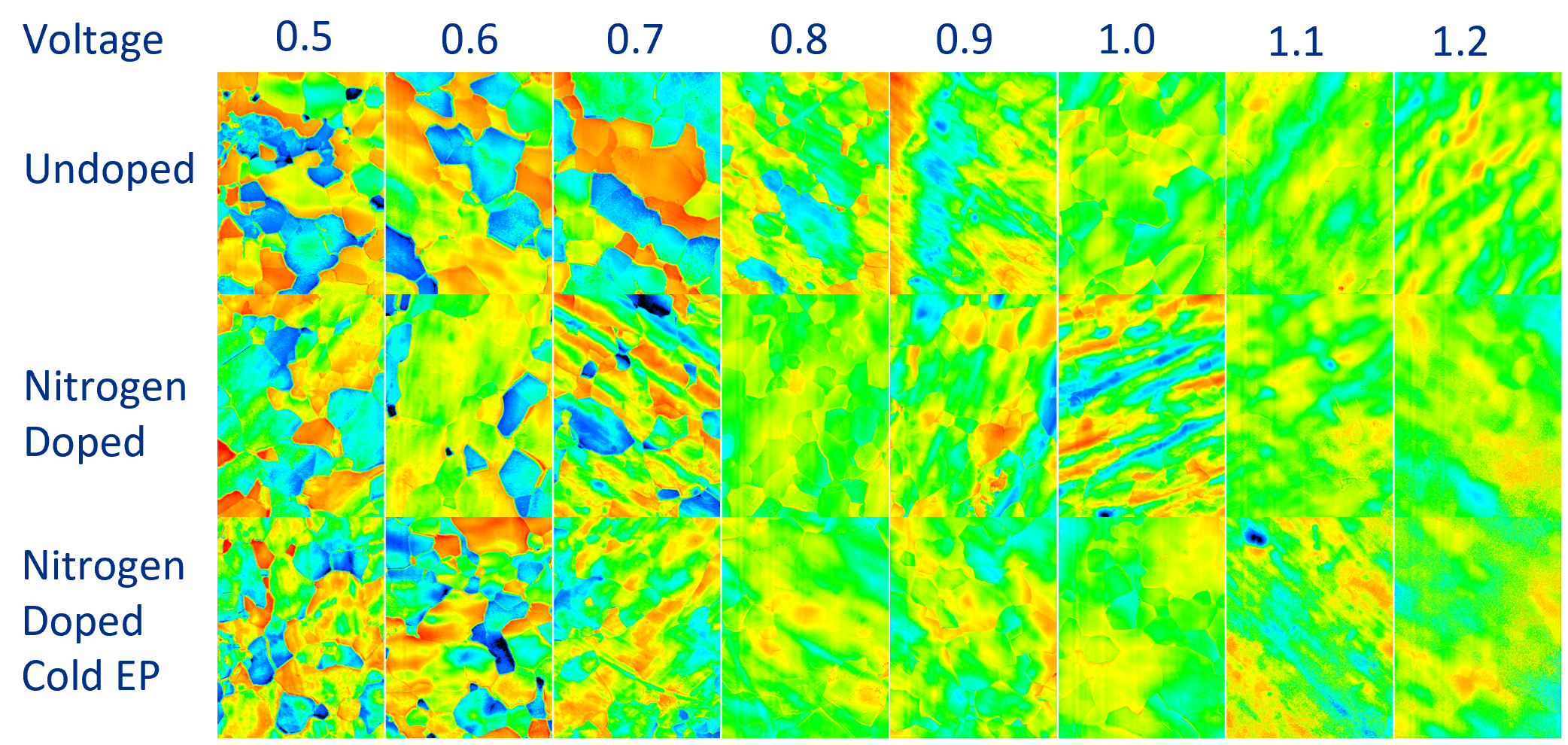}
  \caption{The surface height maps of Nb samples electropolished at different voltages and temperatures.}
  \label{fig:surface_maps}
\end{figure}

Due to slight differences in the oxidation reaction activation energy across the surface, caused by differing grain orientations, surface defects, and the local curvature, the oxide forms preferentially in certain areas. The surface has a mixed oxide with a certain fraction covered by Nb\textsubscript{2}O\textsubscript{5}, NbO, and Nb. As the surface potential is increased a larger portion of the electrode is covered by Nb\textsubscript{2}O\textsubscript{5}. Once the entire surface is covered, the dissolution rate of the surface becomes homogeneous and polishing can occur. If the polishing voltage is not sufficiently high across the surface of the niobium, some areas will experience etching instead of polishing. This effect can be seen in \ref{fig:surface_maps}. On samples polished at low voltage, some of the grains have a facetted surface indicating that grain orientation effects are significant. This faceting disappears when the polishing voltage is above ~\qty{0.8}{\volt}. This effect coincides with the appearance of the Nb\textsubscript{2}O\textsubscript{5} oxidation peak in figure~\ref{fig:gamma}.

Comparing figure \ref{fig:gamma} to the electropolished samples shown in figure \ref{fig:surface_maps} we can see that the appearance of the third peak coincides with a significant change in the surface of the electropolished niobium. The surface goes from an etched surface with facets and large grain boundary steps to a much more polished finish. We theorize that this transition is caused by the stabilization of the oxide film which occurs when the HF is depleted on the niobium electrode surface and the surface oxide transitions from a thin non-homogeneous layer to a thick homogeneous one. Since the oxide film is homogeneous, the effects of grain orientation on the dissolution reaction are eliminated. The result is a smoothing effect on the niobium. At polishing voltages below this critical voltage, the oxide film is too thin or does not cover the entire niobium surface, which leads to etching.

Compared to the small scale samples shown in this study, it is much more difficult to control the polishing voltage when eletcropolishing cavities. This is due to the much larger current and complex geometry of the cavity and cathode. Currently, cavity electropolishing machines do not employ a reference electrode and instead only set a fixed voltage between the anode and cathode. This means that cathode losses and ohmic losses in the electrolyte can significantly alter the polishing potential and even create a non-homogeneous potential distribution over the surface of the cavity. This is why etching can occur even when applying much higher potentials typically used during cavity EP.

To combat the uncertainty of electropolishing cavities it may be possible to apply EIS measurements in the production environment. This would allow for a live diagnostic of the polishing conditions inside the cavity without interfering with the regular EP workflow. This information can be used to tweak the EP parameters such as voltage and temperature to ensure that the cavity is in an optimal polishing regime. This technique may be useful when electropolishing complex cavity geometries such as quarter-wave and half-wave geometries.

\section{Conclusion}
\label{sec:org57282ed}
In this study we find evidence that the etching phenomenon seen in electropolished SRF cavities is caused by insufficient polishing voltage and investigate the mechanism behind it. Niobium polished below a critical voltage will experience etching whereas niobium polished above this voltage will experience smoothing. We theorize that the cause of this change in surface finish is a transition from a partially formed niobium oxide film to a fully formed, stable oxide film. This theory is evidenced by measurements of the niobium electropolishing impedance spectrum over a range of polishing voltages. The impedance spectrum indicates the presence of a blocking mechanism that limits the current. This blocking mechanism is most likely caused by the oxide film which forms when HF is depleted near the niobium surface and the oxidation rate exceeds the rate of dissolution by HF.

\section{Acknowledgements}

The authors would like to thank Dr. Noam Eliaz and Dr. Eyal Sabatani for their insightful discussions and recommendations during the authoring of this manuscript.

This work was produced by Fermi Research Alliance, LLC under Contract No. DE-AC02-07CH11359 with the U.S. Department of Energy, Office of Science, Office of High Energy Physics. Publisher acknowledges the U.S. Government license to provide public access under the DOE Public Access Plan.

\section{Supplemental Information}
\label{sec:sup}

This Supplemental section will cover the mathematical derivations of the equations and numerical techniques used in this paper including the derivation of the first order and higher order impedance response of linear systems and the derivation of the generalized DRT method. The details of calculating the DRT of Nb EP to produce figure~\ref{fig:gamma}, such as chosing the basis functions and normalization parameters, is also shown.

\subsection{The Impedance of a Linearized Dynamical System}

Generally, an electrochemical system such as EP can be described by a state vector $\vec{P}$, which describes all the macroscopic degrees of freedom of the system, and a set of non-linear functions $\vec{g}$, which describes the evolution of each degree of freedom in the system over time. This system is controlled by an independant parameter $E$, usually the electric potential applied to the electrode, and has a measure $I$, usually the current passing through the electrode, which is a non-linear function of $E$ and $\vec{P}$.\cite{wu1998investigation, wu1999general}

\begin{flalign}
    \frac{d\vec{P}}{dt} &= \vec{g}\left(\vec{P},E\right)\\
    I &= f\left(\vec{P},E\right)
\end{flalign}

For a system that satisfies the above mentioned stability condition, the non-linear system of equations can be linearized near its stable point using a first order Taylor expansion, written in matrix notation as shown below.

\begin{flalign}
    \vec{P} &= \vec{P_0} + \delta\vec{P}\\
    E &= E_0 + \delta E\\
    I &= I_0 + \delta I\\
    \frac{d\delta\vec{P}}{dt} &= \mathbf{A}\delta\vec{P} + \mathbf{B}\delta E\\
    \delta I &= \mathbf{C}\delta\vec{P} + \mathbf{D}\delta E\\
    \left(\mathbf{A}\right)_{i,j} &\equiv A_{i,j} = \left(\frac{\partial g_i}{\partial P_j}\right)_{E,P_{k\neq j}}\\
    \left(\mathbf{B}\right)_{i} &\equiv B_i = \left(\frac{\partial g_i}{\partial E}\right)_{\vec{P}}\\
    \left(\mathbf{C}\right)_{i} &\equiv C_i = \left(\frac{\partial f}{\partial P_i}\right)_{P_{k\neq i}}\\
    D &= \left(\frac{\partial f}{\partial E}\right)_{\vec{P}}
\end{flalign}

The matrix $\mathbf{A}$ is the most important parameter as it describes the rate of change of the system's degrees of freedom. We refer to this matrix as the connection matrix due to it's similarity to a kinetic scheme. To calculate the impedance of this linearized system we use the Laplace transformation, $\mathscr{L}$. The impedance is defined as the ratio of the current to the potential in the Laplace domain.

\begin{flalign}
    \widetilde{X} &\equiv \mathscr{L}\left(X\right)\\
    s\delta\widetilde{P} &= \mathbf{A}\delta\widetilde{P} + \mathbf{B}\delta\widetilde{E}\\
    \delta\widetilde{P} &= \left(s\mathbf{I}-A\right)^{-1}\mathbf{B}\delta\widetilde{E}\\
    \delta\widetilde{I} &= \mathbf{C}\delta\widetilde{P} + \mathbf{D}\delta\widetilde{E}\\
    \delta\widetilde{I} &= \mathbf{C}\left(s\mathbf{I}-\mathbf{A}\right)^{-1}\mathbf{B}\delta\widetilde{E} + \mathbf{D}\delta\widetilde{E}\\
    Z\left(s\right) &\equiv \frac{\delta\widetilde{I}}{\delta\widetilde{E}} = \mathbf{C}\left(s\mathbf{I}-\mathbf{A}\right)^{-1}\mathbf{B} + \mathbf{D}
\end{flalign}

In the case that the matrix $\mathbf{A}$ is diagonalizable, finding the inverse of $s\mathbf{I}-\mathbf{A}$ is a case of matrix diagonalization. Let $\mathbf{\Lambda}$ be the diagonal matrix similar to $\mathbf{A}$.

\begin{flalign}
    \mathbf{A} &= \mathbf{P\Lambda P^{-1}}\\
    \left(s\mathbf{I}-\mathbf{A}\right)^{-1} &= \left(s\mathbf{I}-\mathbf{P\Lambda P^{-1}}\right)^{-1}\\
    &= \left(\mathbf{P}\left(s\mathbf{I}-\mathbf{\Lambda}\right)\mathbf{P}^{-1}\right)^{-1}\\
    &= \mathbf{P}\left(s\mathbf{I}-\mathbf{\Lambda}\right)^{-1}\mathbf{P}^{-1}
\end{flalign}

Pluging this expression into the impedance equation, we can see that the impedance of a system with a diagonalizable connection matrix takes the form of a sum of voigt-like terms, a circuit containing a resistor and capacitor. 

\begin{flalign}
    Z\left(s\right) &= \mathbf{C}\mathbf{P}\left(s\mathbf{I}-\mathbf{\Lambda}\right)^{-1}\mathbf{P}^{-1}\mathbf{B} + \mathbf{D}\\
    Z\left(s\right) &= R_s + \sum_{i}\frac{\Omega_i}{s-\lambda_i}\label{eq:first_order_impedance}\\
    \Omega_i &= \left(\mathbf{CP}\right)_i * \left(\mathbf{P}^{-1}\mathbf{B}\right)_i\\
    R_s &= \mathbf{D}
\end{flalign}

Simply set $\lambda = -\frac{1}{RC}$ and $\Omega = \frac{1}{C}$. We also see that the eigenvalues of the system correspond to the relaxation time of an equivalent circuit with resistance $R$ and capacitance $C$.  However, the similarity to a circuit model is purely mathematical and should not be used to infer any physical meaning. In fact, there are multiple circuit combinations that could, with the correct choice of inductors, capacitors, and resistors, achieve the same impedance characteristics.

From this derivation it is clear that the EIS measurement provides information about the eigenvalues of the underlying system dynamics. If the impedance data can be fit to a model consisting of voigt elements, we can extract the eigenvalues, and thus the relaxation times, of the underlying chemical processes. The magnitudes of the $\Omega_i$s indicate the coupling strength of the processes to the input signal. However, the information that can be gained from EIS is limited. This is because the eigenvalues themselves do not define a unique differential equation defined by the matrix $\mathbf{A}$. There are an infinite set of similar matrices that all share the same eigenvalues. Therefore, EIS, on its own, cannot be used to determine the reaction mechanism of an arbitrary electrochemical system. EIS can be used to place limits on a proposed reaction mechanism, or provide information about an already known reaction mechanism.

\subsection{Impedance Characteristics of Systems With Non-Diagonalizable Connection Matrices}

In the previous section, we show that the impedance of a linear system with a diagonalizable connection matrix takes the form of a sum of voigt elements. However, this is not the case in general. If the matrix is not diagonalizable, there will be higher order terms in the impedance. Since the matrix is not diagonalizable, we must instead decompose it into its Jordan canonical form, which is an upper triangular matrix consisting of so-called Jordan blocks along it's diagonal. The Jordan normal form of a matrix and it's pseudo eigen vectors is written like

\begin{flalign}
    \mathbf{A} &= \mathbf{PJP}^{-1}\\
    \mathbf{J} &= diag\left(\mathbf{J}_{\lambda_1,n_1},\ldots,\mathbf{J}_{\lambda_r,n_r}\right)\\
    \mathbf{J}_{\lambda_i,n_i} &= \lambda_i\mathbf{I}+\mathbf{N}_{n_i}
\end{flalign}

Here $\mathbf{N}_{n_i}$ is the upper shift matrix of dimension $n_i$, a matrix containing all zeros except for ones on the first superdiagonal. The inverse of a Jordan matrix is an upper triangular matrix consisting of the inverses of its Jordan blocks.

\begin{flalign}
    \mathbf{J}^{-1} &= diag\left(\mathbf{J}_{\lambda_1,n_1}^{-1},\ldots,\mathbf{J}_{\lambda_r,n_r}^{-1}\right)\\
    \mathbf{J}_{\lambda,n}^{-1} &= \sum_{k=0}^{n-1}\frac{\left(-\mathbf{N}\right)^k}{\lambda^{k+1}}
\end{flalign}

The impedance contribution from one jordan block of size $n_i$ with eigenvalue $\lambda_i$ is

\begin{flalign}
    Z_i\left(s\right) &= \mathbf{C}\mathbf{P}\left(s\mathbf{I}-\mathbf{J}_{\lambda_i,n_i}\right)^{-1}\mathbf{P}^{-1}\mathbf{B}\\
    Z_i\left(s\right) &= \sum_{k=0}^{n_i-1}\mathbf{C}\mathbf{P}\frac{\left(-\mathbf{N}\right)^k}{\left(s-\lambda_i\right)^{k+1}}\mathbf{P}^{-1}\mathbf{B}\\
    Z_i\left(s\right) &= \sum_{k=0}^{n_i-1}\frac{\Omega_{i,k}}{\left(s-\lambda_i\right)^{k+1}}\\
    \Omega_{i,k} &= \sum_{j=0}^{n_i-k-1}\left(-1\right)^k\left(\mathbf{CP}\right)_j * \left(\mathbf{P}^{-1}\mathbf{B}\right)_{j+k}
\end{flalign}

Then the total impedance is then the sum of the impedance contribution from each Jordan block.

\begin{flalign}
    Z\left(s\right) &= \sum_{i=0}^{r}\sum_{k=0}^{n_i-1}\frac{\Omega_{i,k}}{\left(s-\lambda_i\right)^{k+1}}
\end{flalign}

\subsection{The Impedance of a System With Imaginary Eigenvalues}

The previous derivation generalizes to imaginary eigenvalues. For a connection matrix containing only real coefficients the imaginary eigenvalues must appear in complex conjugate pairs. The impedance of an imaginary eigenvalue pair $\lambda_i = \alpha_i \pm j\beta_i$ using equation \ref{eq:first_order_impedance} is

\begin{flalign}
    Z_i\left(s\right) &= \frac{\Omega_+}{s-\alpha_i-j\beta_i} + \frac{\Omega_-}{s-\alpha_i+j\beta_i}\\
    &= \frac{\Omega_+\left(s-\alpha_i+j\beta_i\right) + \Omega_-\left(s-\alpha_i-j\beta_i\right)}{\left(s-\alpha_i-j\beta_i\right)\left(s-\alpha_i+j\beta_i\right)}\\
    &= \frac{\left(\Omega_++\Omega_-\right)\left(s-\alpha_i\right) + \left(\Omega_+-\Omega_-\right)j\beta_i}{\left(s-\alpha_i\right)^2+\beta_i^2}\\
    &= \frac{\Omega_{re}\left(s-\alpha_i\right)+\Omega_{im}j\beta_i}{s^2-2s\alpha_i+\alpha_i^2+\beta_i^2}
\end{flalign}

This impedance response is equivalent to an RLC circuit, a circuit containing a resistor, capacitor, and inductor connected in parallel.

\subsection{Discretizing the Integral of the Generalized DRT Function in Log Coordinates}

It is necessary to discretize the DRT in the log based coordinate system. It is typical to measure the impedance at frequencies spaced logarithmically to cover a wide frequency range. Therefore, it is also necessary to calculate the DRT in logarithmic coordinates. Otherwise, the computational accuracy will be too low at the low frequencies, or too high at the high frequencies. However, this is challenging when using the generalized DRT model, since the integral over $\omega_0$ in equation~\ref{eq:gDRT} spans both positive and negative values. To solve this problem we must separate the integral into the sum of two integrals each integrating over one half of the real axis. After a change of variables and reversing the direction of integration we can substitute $G\left(\omega_0\right)$ with a piece wise fit function $\gamma_{\pm}\left(\ln\omega_0\right)$.

\begin{flalign}
  Z_{gDRT} =& R_{s} + \int_{-\infty}^{0}\frac{G(\omega_0)d\omega_0}{1 + j \frac{\omega}{\omega_0}} + \int_{0}^{\infty}\frac{G(\omega_0)d\omega_0}{1 + j \frac{\omega}{\omega_0}}\\
  Z_{gDRT} =& R_{s} + \int_{0}^{\infty}\frac{G(\omega_0)d\omega_0}{1 + j \frac{\omega}{\omega_0}} + \frac{G(-\omega_0)d\omega_0}{1 - j \frac{\omega}{\omega_0}}\\
  G\left(\omega_0\right) d\omega_0 =& \begin{cases}
    \gamma_+\left(\ln\omega_0\right) d\ln\omega_0 & \omega_0 \ge 0 \\
    \gamma_-\left(\ln-\omega_0\right) d\ln\omega_0 & \omega_0 \le 0 \\
  \end{cases}\\
  Z_{gDRT} =& R_{s} + \int_{-\infty}^{\infty}\frac{\gamma_+\left(\ln\omega_0\right) d\ln\omega_0}{1 + j \frac{\omega}{\omega_0}} + \frac{\gamma_-\left(\ln\omega_0\right) d\ln\omega_0}{1 - j \frac{\omega}{\omega_0}}
\end{flalign}

To find the functions \(\gamma_{\pm}(\ln\omega_0)\) numerically we approximate them using a set of basis functions $\phi_n\left(\ln\omega_0\right)$.

\begin{flalign}
  \gamma_{\pm}(\ln\omega_0)&\approx\sum_{n=1}^{N}x^{\pm}_{n}\phi_{n}(\ln\omega_0)\\
  Z_{gDRT} =& R_{t} + \int_{-\infty}^{\infty}\sum_{n=1}^{N}\frac{x^{+}_{n}\phi_{n}(\ln\omega_0)d\ln\omega_0}{1 + j \frac{\omega}{\omega_0}} + \sum_{n=1}^{N}\frac{x^{-}_{n}\phi_{n}(\ln\omega_0)d\ln\omega_0}{1 - j \frac{\omega}{\omega_0}}
\end{flalign}

Using algebraic manipulation and a change of variables, the integral can be separated into a real part and an imaginary part.

\begin{flalign}
  x_n^{re} =& x_n^+ + x_n^-\\
  x_n^{im} =& x_n^+ - x_n^-\\
  Z_{gDRT} =& R_{s} + \sum_{n=1}^{N}x^{re}_{n} \int_{-\infty}^{\infty} \frac{\phi_{n}(\ln\omega_0)d\ln\omega_0}{1 + \left(\frac{\omega}{\omega_0}\right)^2} + j \sum_{n=1}^{N}x^{im}_{n} \int_{-\infty}^{\infty} \frac{\frac{\omega}{\omega_0}\phi_{n}(\ln\omega_0)d\ln\omega_0}{1 + \left(\frac{\omega}{\omega_0}\right)^2}
\end{flalign}

The impedance at a frequency $\omega = \omega_m$ is a linear function of the values of $x^{re}_n$ and $x^{im}_n$.

\begin{flalign} 
  Z_{gDRT}\left(j \omega_m\right) =& R_{s} + \sum_{n=1}^{N} x^{re}_{n} A^{re}_{n,m} + \sum_{n=1}^{N} x^{im}_{n} A^{im}_{n,m}\\
  A^{re}_{n,m} =& \int_{-\infty}^{\infty} \frac{\phi_{n}(\ln\omega_0)d\ln\omega_0}{1 + \left(\frac{\omega_m}{\omega_0}\right)^2}\label{eq:A_re}\\
  A^{im}_{n,m} =& \int_{-\infty}^{\infty} \frac{\frac{\omega_m}{\omega_0}\phi_{n}(\ln\omega_0)d\ln\omega_0}{1 + \left(\frac{\omega_m}{\omega_0}\right)^2}\label{eq:A_im}
\end{flalign}

Given a set of M experimentally measured impedance data points, $\mathbf{Z_{exp}} = \left[Z_1, \ldots, Z_M\right]$ at frequencies $\mathbf{\omega} = \left[\omega_1, \ldots, \omega_M\right]$ we can find the values of $x^{re}_n$ and $x^{im}_n$ using linear regression.

\begin{flalign}
  \min_{\mathbf{x^{re}},\mathbf{x^{im}}}\lVert\mathbf{Z_{gDRT}}-\mathbf{Z_{exp}}\rVert = \min_{\mathbf{x^{re}},\mathbf{x^{im}}}\left(\lVert \mathbf{x^{re}} \mathbf{A^{re}} - Re\left(\mathbf{Z_{exp}}\right) \rVert + \lVert \mathbf{x^{im}} \mathbf{A^{im}} - Im\left(\mathbf{Z_{exp}}\right) \rVert \right)\label{eq:Zmatrix}
\end{flalign}

\subsection{Normalized Linear Fitting}

When finding the DRT function from experimental data, a normalization method is required to prevent overfitting. In this study we use a modified Tikhonov regularization, based on the work of Wan, et. al.~\cite{wan2015influence}, that penalizes the derivatives of the DRT function. This has the effect of smoothing the function and eliminating oscillations caused by overfitting. With the regularization term added the objective becomes to minimize the residuals and the normalization term together. 

\begin{flalign}
  \min_{\mathbf{x^{re}},\mathbf{x^{im}}}\left(\lVert \mathbf{x^{re}} \mathbf{A^{re}} - Re\left(\mathbf{Z_{exp}}\right) \rVert + \lVert \mathbf{x^{im}} \mathbf{A^{im}} - Im\left(\mathbf{Z_{exp}}\right) \rVert + \mathbf{x^{re} M x^{re}}^T + \mathbf{x^{im} M x^{im}}^T \right)
\end{flalign}

The matrix $\mathbf{M}$ is calculated by integrating the square of the derivatives of the test functions, $\phi_n\left(\omega_0\right)$. Calculating the zeroth derivative, i.e. the function itself, is equivalent to standard Tikhonov regularization. The strength of the regularization is controlled by multiplying by a constant $\lambda_k$, where $k$ is the k-th derivative of the test functions.

\begin{flalign}
  (\mathbf{M}_{k})_{n,m} =& \int_{-\infty}^{\infty} \frac{d^{k}\phi_{n}}{dln\omega^{k}} \frac{d^{k}\phi_{m}}{dln\omega^{k}} dln\omega\\
  \mathbf{M} =& \sum_{k=0}^{K}\lambda_{k}\mathbf{M}_{k}
\end{flalign}

The optimum values of \(\lambda_k\) are difficult to find mathematically, so values were manually adjusted up to $k=2$ to eliminate oscillations without sacrificing the accuracy of the fit. Mathematical heuristics such as the L-curve, cross-validation, Fourier transform\cite{BOUKAMP201712} or Bayesian methods\cite{ciucci2015analysis} can be used to more precisely optimize the regularization parameters.

\subsection{Basis Function Selection}
\label{sec:org8198a5a}

The choice of basis function is quite arbitrary as long as the resulting function space is large enough to approximate the exact solution for the DRT function. However, in practice the basis function has a large impact on the number of parameters required and the level of regularization required. The basis function should also be relatively easy to compute using a computer to speed up the computation and the function should be differentiable so that the regularization matrix can be calculated. A natural choice is to use the log Gaussian, which has been shown to be effective at fitting the DRT function for several kinds of impedance systems\cite{wan2015influence}. These functions are gaussian functions in the log coordinates shifted by $\ln\left(\omega_n\right)$.

\begin{flalign}
  \phi_{n}(ln\omega) &= \frac{1}{\sigma\sqrt{2\pi}}e^{-\left(\frac{ln\omega-ln\omega_{n}}{\sigma\sqrt{2}}\right)^2}
\end{flalign}

Combining this test function with equation~\ref{eq:A_re} and \ref{eq:A_im} gives the following expressions for the matrices $A_{re}$ and $A_{im}$.

\begin{flalign}
  A^{re}_{n,m} =& \frac{1}{\sigma\sqrt{2\pi}} \int_{-\infty}^{\infty} \frac{e^{-\left(\frac{ln\omega_0-ln\omega_{n}}{\sigma\sqrt{2}}\right)^2} d\ln\omega_0}{1 + e^{2\left(\ln\omega_m - \ln\omega_0\right)}}\\
  A^{im}_{n,m} =& \frac{1}{\sigma\sqrt{2\pi}} \int_{-\infty}^{\infty} \frac{e^{\ln\omega_m-\ln\omega_0} e^{-\left(\frac{ln\omega_0-ln\omega_{n}}{\sigma\sqrt{2}}\right)^2} d\ln\omega_0}{1 + e^{2\left(\ln\omega_m - \ln\omega_0\right)}}
\end{flalign}

These integrals are calculated numerically and used to find the DRT.

\printbibliography

\end{document}